\documentclass[12pt]{article}
\usepackage{amssymb}\usepackage{amsmath}\usepackage{amsthm}

\newcommand{\be}{\begin{equation}}
\newcommand{\ee}{\end{equation}}
\newcommand{\bea}{\begin{eqnarray}}
\newcommand{\eea}{\end{eqnarray}}
\newcommand{\bg}{\begin{gather}}
\newcommand{\eg}{\end{gather}}

\newcommand{\RR}{\mathbb R}
\newcommand{\ZZ}{\mathbb Z}
\newcommand{\NN}{\mathbb N}

\newcommand{\supp}{{\rm supp\>}}

\begin{document}
%%%%%%%%%%%%%%%%%%%%%%%%%%%%%%
\title{Polymer state approximation of Schr\"odinger wave functions}
\author{
Klaus Fredenhagen and Felix Reszewski \\[2mm]
II. Institut f\"ur Theoretische Physik\\
%Universit\"at Hamburg\\
%Luruper Chaussee 149\\
D-22761 Hamburg, Germany\\
{\tt \small klaus.fredenhagen@desy.de}}
%%%%%%%%%%%%%%%%%%%%%%%%%%%%%%%%%%%%%%%%%%%%%%%%%%%%%%%%%%%%%%%%%%%

%\date{}
\maketitle
\begin{abstract}
It is shown how states of a quantum mechanical particle in the Schr\"odinger representation can be approximated by states in the so-called polymer representation. The result may shed some light on the semiclassical limit of loop quantum gravity.
%{\bf PACS.} 

\end{abstract}

%\tableofcontents
%%%%%%%%%%%%%%%%%%%%%%%%%%%%%%%%%%%%%%%%%%%%%%%%%%%%%%%%%%%%%%%%%%%%%%%%%%%%
\section{Introduction}\setcounter{equation}{0}
%%%%%%%%%%%%%%%%%%%%%%%%%%%%%%%%%%%%%%%%%%%%%%%%%%%
A fundamental feature of the algebraic formulation of quantum physics is the fact, that the states of any faithful representation of a C*-algebra form a $*$-weakly dense subset of the full state space of the algebra (Fell' s Theorem)(see e.g. \cite{Haag,HK}). This general fact becomes relevant if one tries to compare states in the
so-called polymer representations of loop quantum gravity with states occuring in quantum field theory. It is a highly debated question whether loop quantum gravity has the potential to describe continuum physics in an appropriate limit (see, e.g. \cite{Helling}-%,Ashtekar04,Thiemann:2002vj,Thiemann:2004qu,Varadarajan:2001nm,
\cite{Varadarajan:1999it}). A toy model for which this question can be discussed is provided by quantum mechanics of a single particle in 1 spatial dimension (see \cite{Ashtekar03,Helling}). This model also has direct relevance for cosmological considerations (see \cite{Ashtekar:2006bp}).

Whereas the critics of the principal possibility to approximate states in the standard Schr\"odinger representation by states in a singular representation of the canonical commutation relations is unjustified in view of Fell's Theorem, the answers in the affirmative given so far (see \cite{Ashtekar03}) are not completely satisfactory. Namely, in this paper, expectation values in some state in the Schr\"odinger representation are approximated by linear functionals obtained by pairing vectors in a dense subspace of the representation space of the polymer representation (called the subspace of cylindrical functions) with elements of the dual. One would like to interpret these functionals as expectation values of state vectors in the polymer representation. But they are neither normal with respect to the polymer representation (e.g. a sequence of Weyl operators may converge weakly to zero
in the polymer representation, but their values in these functionals may approach a finite value) nor are they positive. 
On the other hand, the answer via Fell's Theorem suffers from the fact that the theorem does not give an explicit construction of the approximating states (it is based on the Hahn-Banach Theorem). We therefore aim in this note at closing these gaps.
%%%%%%%%%%%%%%%%%%%%%%%%%%%%%
\section{Weyl algebra and polymer representation}\label{weyl}
In the Schr\"odinger representation, the Weyl operators
\be W(\alpha,\beta):=e^{i(\alpha q +\beta p)}\ee
with the standard momentum and position operators $p$ and $q$ and $\alpha,\beta\in\RR$, satisfy the Weyl relation 
\be W(\alpha_1,\beta_1)W(\alpha_2,\beta_2)=e^{-\frac{i}{2}(\alpha_1\beta_2-\alpha_2\beta_1)}W(\alpha_1+\alpha_2,\beta_1+\beta_2) \ .\ee
Together with the unitarity condition
\be W(\alpha,\beta)^*=W(-\alpha,-\beta)\ee
these relations alone define a unique simple C*-algebra, the Weyl algebra. The Schr\"odinger representation is (up to unitary equivalence) the only irreducible representation of the Weyl algebra in which the Weyl operators are continuous functions of $\alpha$ and $\beta$ (with respect to the weak operator topology). 

There are many irreducible
representations where this continuity condition is not satisfied. One special example is the so-called polymer representation. The Hilbert space $\mathfrak{H}_{\text{poly}}$ in this representation consists of functions $\Psi$ on the real line which vanish up to a countable subset and satisfy the condition 
\be \sum_{x\in\RR}|\Psi(x)|^2<\infty \ ,\ee
and the scalar product is defined by 
\be (\Psi,\Phi)=\sum_{x\in\RR}\overline{\Psi(x)}\Phi(x) \ .\ee
The Weyl operators act on these functions in the same way as in the Schr\"odinger representation,\footnote{To keep the notation simple we will use the same symbol for a Weyl operator in both representations. There will be no ambiguities since we denote state vectors in the Schr\"odinger Hilbert space by lower case greek letters and state vectors in $\mathfrak{H}_{\text{poly}}$ by capital greek letters.}
\be (W(\alpha,\beta)\Psi)(x)=e^{-\frac{i}{2}\alpha\beta}e^{i\alpha x}\Psi(x-\beta) \ .\ee
  
The position operator may be defined as usual on a dense subspace and possesses even a complete set of (normalizable) eigenvectors $\{|x\rangle,x\in\RR\}$; the momentum operator, however, cannot be defined.  

This well known representation shares some similarities with the so-called polymer representations in Loop Quantum Gravity  and may serve as a toy model for the discussion of structural problems. In \cite{Ashtekar03} the question was discussed in which way states in the Schr\"odinger representation can be approximated from the polymer representation. An inductive system of countable subsets $M$ of $\RR$ was found which has the property that for Schwartz space functions $\psi$ the restriction to any of these subsets $M$ defines an element $P_M\psi$ of the polymer Hilbert space. 
Every such wave function defines a linear functional on the inductive limit of the corresponding subspaces, called the space of cylindrical functions. We will denote this functional by $\left\langle\psi\right|$ and write its action on a state vector $\Phi$ as $\left\langle\psi|\Phi\right\rangle$. This action is defined by
\be \left\langle\psi|\Phi\right\rangle = (P_S\psi,\Phi) \ , \qquad \mathrm{where} \qquad S=\supp\Phi.\ee
In order to define expectation values, the set $M$ was specified to be a lattice of the form $\varepsilon\ZZ$. The expectation value in the state $\psi$ was then approximated by
\be \varepsilon\left\langle\psi|AP_M\psi\right\rangle \ ,\ee 
where $A$ is a finite linear combination of Weyl operators. Clearly, in the limit $\varepsilon\to 0$ the above expression converges to the expectation value in the Schr\"odinger representation. 

Unfortunately, as already mentioned in the introduction, the approximation above
cannot be understood as an approximation of Schr\"odinger states by polymer states in the sense of expectation values. 
First of all, the linear functional 
\be A\to  \varepsilon\left\langle\psi|AP_M\psi\right\rangle\ee
is not normal with respect to the polymer representation, hence cannot be described in terms of matrix elements in this representation. Namely, consider the sequence $W(0,\frac{1}{n})$, $n\in\NN$. Its matrix elements between arbitrary position eigenstates tend to zero, hence, being a bounded sequence, it will converge to zero in the weak operator topology. On the other hand, in the linear functional above, we find
\be\varepsilon\sum_{z\in\ZZ}\overline{\psi(\varepsilon z+\tfrac{1}{n})}\psi({\varepsilon z})\to \varepsilon\sum_{z\in\ZZ}|\psi(\varepsilon z)|^2  \ .\ee
The second problem is, that these functionals are not positive.   
Namely, choose 
\be\psi(x)=e^{i\beta x}e^{-\frac{x^2}{2}}\ee 
as the Schr\"odinger wave function to be approximated. Choose $M=\varepsilon (\mathbb{Z}+\lambda)$ as a countable subset of the real line and the so-called shadow state 
\be P_M \psi=\sum_{x\in M}\psi(x)|x\rangle\ee
We compute the approximate expectation value of the positive operator 
\be A=(1-V(\alpha))^*(1-V(\alpha))=2-V(\alpha)-V(-\alpha) \ .\ee
where $V(\alpha) = W(0,\alpha)$. We obtain
\be \varepsilon\langle \psi|(2-(V(\alpha)+V(-\alpha))P_M \psi\rangle = \ee
\be \varepsilon\sum_{x\in M} (2|\psi(x)|^2-
(\overline{\psi(x+\alpha)+\psi(x-\alpha)})\psi(x))=\ee
\be \varepsilon \sum_{x\in M} 2e^{-x^2}(1-(\cosh \alpha x\cos\alpha\beta-i\sinh\alpha x\sin\alpha\beta )e^{-\frac{\alpha^2}{2}})\ee        
which, in general, is not real.
%%%%%%%%%%%%%%%%%%%%%%%%%%%%%%
\section{Approximation by polymer states}
%%%%%%%%%%%%%%%%%%%%%%%%%%%%%%
Let $\psi\in L^2(\RR)$ be a normalized wave function in the Schr\"odinger representation. Let $A=(A_1,\ldots,A_n)$ be a finite number of elements of the Weyl algebra and let $\varepsilon=(\varepsilon_1,\ldots,\varepsilon_n)$ be a family of positive numbers. 
We search for a unit vector $\Psi$ in the polymer representation such that 
\be |(\psi,A_i\psi)-(\Psi,A_i\Psi)|<\varepsilon_i\ ,\ i=1,\ldots,n \ee
We may find Weyl operators $W(\alpha_{ik},\beta_{ik})$, $k=1,\ldots N_i$, complex numbers $\lambda_{ik}$, such that
\be ||A_i-\sum_k\lambda_{ik}W(\alpha_{ik},\beta_{ik})||<\frac{\varepsilon_i}{3}\ee
We therefore may look for a vector $\Psi$,
such that
\be |(\psi,W(\alpha_{ik},\beta_{ik})\psi)-(\Psi,W(\alpha_{ik},\beta_{ik})\Psi)|<\delta_{ik}\ee
with $\sum_k |\lambda_{ik}|\delta_{ik}<\frac{\varepsilon_i}{3}$. 

The expectation value in the Schr\"odinger representation has the form
\be (\psi,W(\alpha,\beta)\psi)=\int dx \, \overline{\psi(x)}e^{i\alpha (x +\frac{1}{2}\beta)}\psi(x-\beta)\ .\ee
In the polymer representation we have instead
\be(\Psi,W(\alpha,\beta)\Psi)=\sum_{x\in\RR} \overline{\Psi(x)}e^{i\alpha (x +\frac{1}{2}\beta)} \Psi(x-\beta)  \ .\ee
Therefore one might try to choose $\Psi$ such that the latter sum is a Riemann approximation to the integral above. 
But for $\Psi$ normalizable, the coefficients $\Psi(x)$ can be different from zero only on a countable subset $\supp\Psi \subset\RR$. Then the sum extends only over $x$ in the intersection $\supp\Psi\cap\supp\Psi +\beta$. 
To ensure that the intersection is sufficiently large, one may choose a countable subset which is invariant under translation by $\beta_n$, $n=1,\ldots,N$. But in the generic case such a set is dense in 
$\RR$, hence the coefficients $\Psi(x)$ can not be identified with the values of the wave function 
$\psi(x)$, multiplied by the square root of the length of an appropriate interval.

Instead we may look at the additive subgroup of $\RR$ which is generated by $\beta_n$, $n=1,\ldots,N$. This subgroup is a torsion free abelian group and therefore isomorphic to $\ZZ^L$ for some $L\le N$. The isomorphism $\Gamma$ may be considered as the projection which maps the lattice $\ZZ^L$ onto a quasilattice in $\RR$. It has a unique extension to a linear map from $\RR^L$ onto $\RR$ which we will denote by $\gamma$. $\gamma$ may be identified with an element of $\RR^L$ such that $\gamma(z)=\sum_i \gamma_i z_i$.
We now choose a function $\chi$ of one real variable which is continuous, has compact support and satisfies the normalization condition
\be \int_{\RR^{L-1}}dz^{L-1}|\chi(|z|^2)|^2 =1 \ee
where $|z|^2=\sum_i z_i^2$.
We approximate $\psi$ within $L^2(\RR)$ by a continuous function $\phi$ with compact support and define a function $\phi_\chi$ on $\RR^L$ by
\be \phi_\chi(z)=|\gamma|^{\frac L2} \phi(\gamma(z)) \chi(|\gamma|^2|z|^2-\gamma(z)^2) \ .\ee
We then define approximating vectors 
$\Psi_{m,\chi}$, $m\in\NN$ in the polymer space by
\be \Psi_{m,\chi}=m^{-\frac{L}{2}}\sum_{z\in \frac{1}{m}\ZZ^L}\phi_\chi(z)|\gamma(z)\rangle \ . \ee
Inserting this into the formula for the expectation value we obtain
\be (\Psi_{m,\chi},W(\alpha_n,\beta_n)\Psi_{m,\chi})=m^{-L}\sum_{z\in \frac{1}{m}\ZZ^L}\overline{\phi_\chi(z)}e^{i\alpha_n (\gamma(z) +\frac{1}{2}\beta_n)}\phi_\chi(z- \Gamma^{-1}(\beta_n))\ .\ee
The latter expression is a Riemann approximation of the corresponding integral and will converge as $m$ tends to infinity. The limit, however, will depend on the choice of $\chi$. We obtain
\begin{eqnarray*} 
\lim_{m\to\infty}(\Psi_{m,\chi},W(\alpha_n,\beta_n)\Psi_{m,\chi}) = \int d^Lz\, \overline{\phi_\chi(z)}e^{i\alpha_n (\gamma(z) +\frac{1}{2}\beta_n)}\phi_\chi(z- \Gamma^{-1}(\beta_n)) \\
= (\phi,W(\alpha_n,\beta_n)\phi)\int_{\gamma(z)=0} d^{L-1}z\, \overline{\chi(|z|^2)}\chi(|z-|\gamma| \Gamma^{-1}(\beta_n)|^2-\beta_n^2) 
\end{eqnarray*}
where, with an appropriate choice of coordinates, we separated the integral in the first line and obtained the expectation value in the state vector $\phi$ multiplied with an $(L\!-\!1)$-dimensional integral over the kernel of the function $\gamma$. Finally, in the last step, we choose $\chi$ such that the integral in the second line approaches unity. We may, e.g., scale $\chi$ by setting 
\be \chi_\lambda(|z|^2)=\lambda^{\frac{L-1}{2}}\chi(\lambda^2 |z|^2) \ ,\ee
get
\begin{eqnarray*}
&&\int_{\gamma(z)=0} d^{L-1}z\, \overline{\chi_\lambda(|z|^2)}\chi_\lambda(|z-|\gamma| \Gamma^{-1}(\beta_n)|^2-\beta_n^2)= \\
&&\int_{\gamma(z)=0} d^{L-1}z\, \overline{\chi(|z|^2)}\chi(|z-\lambda|\gamma| \Gamma^{-1}(\beta_n)|^2-\lambda^2\beta_n^2)
\end{eqnarray*}
and perform the limit $\lambda\to0$.

\subsection{Example}

As an example of this method consider the simple case where the additive subgroup of $\mathbb R$ is generated by $\beta_1 = 1$ and $\beta_2 = \sqrt 2$. The map $\Gamma(z) = \sum_i \beta_i z_i$ is then already an isomorphism. For the function $\phi_\chi$ one gets
\be \phi_\chi(z) = \sqrt3 \, \phi(z_1 + \sqrt2\, z_2)\, \chi(2z_1^2 + z_2^2 - 2\sqrt2\, z_1z_2) \ . \ee
One can introduce new coordinates $z_1',z_2'$ defined by
\begin{eqnarray}
z_1' &=& z_1 + \sqrt2 \, z_2 \ , \nonumber\\
z_2' &=& \sqrt2 \, z_1 - z_2 \ .
\end{eqnarray}
\parbox[t]{5.5cm}{\vspace{0.5cm}This corresponds to a rotation of the coordinate system such that the $z_1'$-axis now points in the direction defined by the vector $\beta = (\beta_1,\beta_2)$ in the old coordinates. The isomorphism $\Gamma$ projects the points of $\ZZ^2$ onto the $z_1'$-axis. This is illustrated in the picture.}
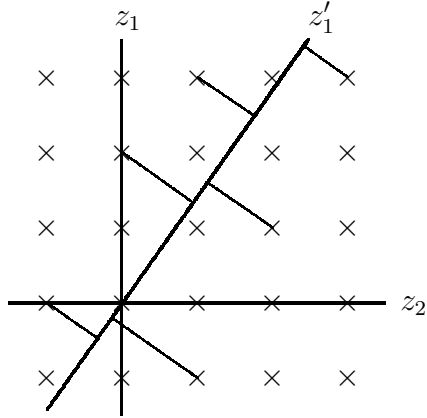
\hfill \parbox[t]{8cm}{
\begin{center}
\setlength{\unitlength}{1cm}
\begin{picture}(5.5,5.5)(-2,-2)
\thicklines
\qbezier(-1,-1.41)(1.73,2.46)(2.46,3.5)
\put(-1.5,0){\line(1,0){5}}
\put(0,-1.5){\line(0,1){5}}
\put(-0.1,3.7){\mbox{$z_1$}}
\put(3.7,-0.1){\mbox{$z_2$}}
\put(2.48,3.7){\mbox{$z_1'$}}
\multiput(-1.165,2.89)(1,0){5}{\mbox{$\times$}}
\multiput(-1.165,1.89)(1,0){5}{\mbox{$\times$}}
\multiput(-1.165,0.89)(1,0){5}{\mbox{$\times$}}
\multiput(-1.165,-0.11)(1,0){5}{\mbox{$\times$}}
\multiput(-1.165,-1.11)(1,0){5}{\mbox{$\times$}}
\thinlines
\qbezier(-1,0)(-0.67,-0.24)(-0.33,-0.47)
\qbezier(1,-1)(0.43,-0.6)(-0.14,-0.2)   
\qbezier(0,2)(0.47,1.67)(0.94,1.33)   
\qbezier(2,1)(1.57,1.3)(1.13,1.6)            
\qbezier(1,3)(1.37,2.75)(1.75,2.49)        
\qbezier(3,3)(2.7,3.21)(2.41,3.42)   
\end{picture}
\end{center}}

With these new coordinates the approximating vector in the polymer space has the form
\be \Psi_{m,\chi} = m^{-1}\, \sum_{z\in\frac1m \ZZ^2}\, \sqrt3 \, \phi(z_1')\, \chi(z_2'^2)\, |z_1'\rangle \ . \ee
One immediately sees that the corresponding integral for the expectation value separates.

\section{Approximation of the momentum operator}

As mentioned in section \ref{weyl}, in the polymer representation a momentum operator cannot be defined. One may ask, in which sense the Schr\"odinger momentum operator can be approximated within the polymer representation. Consider, for instance, the coherent Schr\"odinger state
\be \psi(x) = (\pi d^2)^{-\frac14} \; \exp \left( -\frac{(x-x_0)^2}{2d^2}+i p_0(x-x_0) \right) \ , \ee 
where $d$ is a length scale, such that the inverse of $d$ is proportional to the uncertainty in $p$. The expectation value for $p$ in this state is $p_0$, the expectation value of $p^2$ is $\langle p^2 \rangle = p_0^2 + \frac{1}{d^2}$ . For $\beta^2\langle p^2 \rangle\ll 1$ 
the operator (see \cite{Ashtekar03})
\be p_\beta = \frac i{2\beta}(V(\beta)-V(-\beta)) \ee
is an approximation of the standard Schr\"odinger momentum operator. In particular, 
the expectation value of $p_\beta$ in the coherent state above is 
\bea(\psi,p_\beta\psi) &=& \frac i{2\beta} \,e^{-\frac{\beta^2}{4d^2}} \left( e^{-i p_0\beta}-e^{i p_0\beta}\right) \nonumber \\
&=& p_0 \left(1+{\cal O}(\langle p^2 \rangle\beta^2)\right) \ . \eea

For a given $\beta$ we may now consider the expectation value of $p_{\beta}$ in the polymer 
state
\be \Psi_\beta = \beta^{\frac12} \sum_{z \in \ZZ} \psi(\beta z) |\beta z\rangle . \ee
and obtain the approximation 
\be \left| (\Psi_\beta,p_\beta\Psi_\beta) - (\psi,p\psi) \right| \sim {\cal O}(p_0\langle p^2\rangle\beta^2) \ . \ee
The difficulty is that for every choice of $\beta$ one has to use a different polymer state. It is impossible to find a polymer state which approximates the expectation value of the momentum for all sufficiently small values of $\beta$. This is due to the fact that the polymer representation is not weakly continuous in the parameter $\beta$. It is however possible, as explicitly shown in this paper, to find approximations for any finite set of $\beta$'s. For instance, let $\beta_1,\beta_2 \ll \langle p^2\rangle ^{\frac12}$ with $\beta_1/\beta_2$ irrational. Then an approximating polymer state is 
\be \Psi_{\beta_1\beta_2}= (\beta_1^2+\beta_2^2)^{\frac12} \sum_{z_1,z_2 \in \ZZ} \psi(\beta_1 z_1 + \beta_2 z_2) \, \chi_\lambda \left((\beta_2z_1-\beta_1z_2)^2\right) \; |\beta_1z_1 + \beta_2z_2\rangle \ ,\ee
where we may choose
\be \chi_\lambda(|z|^2) = \sqrt{\frac\lambda\pi} \, e^{-\lambda^2|z|^2} \ee
with $\lambda < \langle p^2\rangle^{\frac12}$. 

%%%%%%%%%%%%%%%%%%%%%%%%%%%
\section{Conclusions}
Given any state of a quantum mechanical particle in the 
Schr\"odinger representation we constructed a net of states in the polymer 
representation of the Weyl algebra such that the expectation values of all 
elements of the Weyl algebra converge pointwise to the expectation values in 
the given state. The existence of such a net follows from Fell's Theorem 
(density of the states of one faithful representation in the set of all states
with respect to the weak-*-topology on the dual of a C*-algebra), but the 
proof of the theorem is not constructive and therefore does not amount to an 
explicit construction. Previous explicit approximations of states in the 
Schr\"odinger representations were in terms of linear functionals which 
could not be interpreted as expectation values of states in the polymer 
representation. 

As a byproduct we proved that pure states in the Schr\"odinger representation 
can be approximated by pure states in the polymer representation. This goes 
beyond the assertion of Fell's Theorem. 

Observables which can only be defined in the Schr\"odinger representation 
(as the momentum operator discussed in the previous section) 
have first to be approximated by linear combinations of Weyl operators 
(in the sense of expectation values and, possibly, uncertainties). For a 
finite number of these approximations one then can find polymer states with 
approximately equal expectation values (and uncertainties). 

It depends on the problem under investigation whether the proven convergence 
is strong enough. Since the representations are inequivalent, a uniform 
approximation is not possible. In particular, the question whether the spectrum
of an observable is discrete or continuous depends on the equivalence class 
of the representation (as exemplified by the position operator). Moreover, it 
is not possible to replace the net of states by a sequence. Namely, given any 
sequence $\Psi_n$ of normalized wave functions in the polymer Hilbert space, 
the expectation value of $W(\alpha,\beta)$ vanishes for all $n$ up to a 
countable set of values for $\beta$. In the Schr\"odinger representation, 
however, the expectation value in any given state must be near to unity for 
small values of $\alpha$ and $\beta$.  

%%%%%%%%%%%%%%%%%%%%%%%
\vskip0.5cm
{\bf Acknowledgements:} We are grateful to Abhay Ashtekar for making available to us the manuscript \cite{Ashtekar06} prior to publication and for useful comments on a preliminary version of this paper.

%\newpage


\begin{thebibliography}{999}
\bibitem{Ashtekar03} Abhay Ashtekar, Stephen Fairhurst and Joshua L. Willis, 
  ``Quantum gravity, shadow states, and quantum mechanics'', 
  Class. Quant. Grav. \textbf{20} (2003) 1031-1062 

\bibitem{Ashtekar06} Abhay Ashtekar, Stephen Fairhurst and Amit Glosh, ``U(1) polymer and fock representations'', in preparation
\bibitem{Haag} Haag, R., 
  ``Local Quantum Physics: Fields, particles and algebras'', 
  Springer-Verlag, Berlin, 2nd ed. (1996)

\bibitem{HK} Haag, R., and Kastler, D., 
  ``An algebraic approach to field theory'', 
  Journ. Math. Phys. \textbf{5} (1964) 848

\bibitem{Helling}
  R. C. Helling and G. Policastro,
  ``String quantization: Fock vs. LQG representations,''
  arXiv:hep-th/0409182.

\bibitem{Ashtekar04}
  A. Ashtekar and J. Lewandowski,
  ``Background independent quantum gravity: A status report,''
  Class.\ Quant.\ Grav.\  {\bf 21} (2004) R53
  [arXiv:gr-qc/0404018].

\bibitem{Thiemann:2002vj}
  T. Thiemann,
  ``Complexifier coherent states for quantum general relativity,''
  Class.\ Quant.\ Grav.\  {\bf 23} (2006) 2063
  [arXiv:gr-qc/0206037].

\bibitem{Thiemann:2004qu}
  T. Thiemann,
  ``The LQG string: Loop quantum gravity quantization of string theory. I:
  Flat target space,''
  Class.\ Quant.\ Grav.\  {\bf 23} (2006) 1923
  [arXiv:hep-th/0401172].

\bibitem{Varadarajan:2001nm}
  M. Varadarajan,
  ``Photons from quantized electric flux representations,''
  Phys.\ Rev.\ D {\bf 64} (2001) 104003
  [arXiv:gr-qc/0104051].

\bibitem{Varadarajan:1999it}
  M. Varadarajan,
  ``Fock representations from U(1) holonomy algebras,''
  Phys.\ Rev.\ D {\bf 61} (2000) 104001
  [arXiv:gr-qc/0001050].

\bibitem{Ashtekar:2006bp}
  A.~Ashtekar,
  ``Gravity, geometry and the quantum,''
  arXiv:gr-qc/0605011.

\end{thebibliography}
\end{document}